# The Community Authorization Service: Status and Future

L. Pearlman, C. Kesselman
*USC Information Sciences Institute, Marina del Rey, CA*

V. Welch, I. Foster, S. Tuecke
*Argonne National Laboratory, Argonne, IL*

Virtual organizations (VOs) are communities of resource providers and users distributed over multiple policy domains. These VOs often wish to define and enforce consistent policies in addition to the policies of their underlying domains. This is challenging, not only because of the problems in distributing the policy to the domains, but also because of the fact that those domains may each have different capabilities for enforcing the policy. The Community Authorization Service (CAS) solves this problem by allowing resource providers to delegate some policy authority to the VO while maintaining ultimate control over their resources. In this paper we describe CAS and our past and current implementations of CAS, and we discuss our plans for CAS-related research.

## 1. INTRODUCTION

A virtual organization (VO) is a dynamic collection of resources and users unified by a common goal and potentially spanning multiple administrative domains [12] VOs introduce challenging management and policy issues, resulting from often complex relationships between local site policies and the goals of the VO with respect to access control, resource allocation, and so forth. In particular, authorization solutions are needed that can empower VOs to set policies concerning how resources assigned to the "community" are used—without, however, compromising site policy requirements.

We describe here our implementation of the Community Authorization Service (CAS), a system that we have developed as part of a solution to this problem. CAS allows for a separation of concerns between site policies and VO policies. Specifically, sites can delegate management of a subset of their policy space to the VO. CAS provides a fine-grained mechanism for a VO to manage these delegated policy spaces, allowing it to express and enforce expressive, consistent policies across resources spanning multiple independent policy domains. Both past and present CAS implementations build on the Globus Toolkit® middleware for Grid computing [10], thus allowing for easy integration of CAS with existing Grid deployments.

The rest of this article is as follows. In Section 2, we describe the scenarios that CAS is designed to support, review the Globus Toolkit's authorization systems, and explain why CAS is needed. In Section 3, we introduce CAS architecture and concepts. In Section 4, we describe the current and present CAS implementations and future implementation plans. In Section 5, we discuss the relationship of CAS to PERMIS, VOMS, and other related authorization technologies and how standardization can enable interoperability among them. In Section 6, we describe our plans for future CAS-related research.

## 2. PROBLEM STATEMENT

We first describe the scenarios that CAS was designed to solve, review the current Globus Toolkit authorization system, and explain the system limitations that motivated development of CAS.

### 2.1. Consistent Distributed Policy

While all the physical organizations in a virtual organization may in principle agree to allow the VO access to its resources, each organization will typically retain ultimate control over the policies that govern access to its resources. Nevertheless, the VO will often wish to apply some common policy about how its users access the resources assigned to the VO. For example, a subset of users may have read-only access to VO data, while others may have full access to publish or modify data. Some data may be sensitive until published and restricted to a very limited set of VO members.

Since VO resources are located within multiple organizations, maintaining a consistent policy means having access control enforced by each of the organizations in a consistent manner. Achieving this consistency is difficult because each site potentially has different mechanisms for policy expression and enforcement and these different mechanisms may have different abilities in terms of the granularity of the policy they can express. Using local policy mechanisms to enforce VO policies would limit those policies to the least common denominator of the abilities of the site mechanisms (which can potentially change when new sites are added to the VO).

Complicating this situation is the fact that the resources may be dynamic and may also have dynamic policies. For example, a VO can have a number of datasets as part of its resources. While these datasets are stored on resources owned by the organizations that the VO spans, the VO wishes to apply its own policy regarding access to the datasets. These datasets may be dynamically created or copied; for example, a demand for access to a dataset by a user at a particular location might motivate a replica of the dataset being automatically created at the location to improve the user's access (as described in [1]). These data replicas should have access control policies identical to the original from they are copied, in order to maintain a consistent policy in regard to access to the data. This policy in turn may be dynamic, meaning the policy on the replicas must mirror it as it changes.

The use of site mechanisms also has the complication that policy itself must be replicated: a change in the policy requires an update at multiple sites. The temporarily unavailability of a site during such an update, can lead to inconsistencies in the policy across the VO.

## 2.2. Authorization Middleware

Middleware for VOs has been available for several years. Most notable is the Globus Toolkit (GT) [10], which is freely available and widely deployed.

### 2.2.1. Globus Toolkit Features

The Globus Toolkit normally uses existing local resource mechanisms for authorization. A user is authenticated and then mapped to a local identity (e.g., a Unix account) by a local configuration file (the *grid-mapfile*). This mapping also serves as an access control check: if the user is not listed in the local mapping configuration, access to the resource is denied.

Once the user is mapped to a local identity, the Globus Toolkit then relies solely on local policy management and enforcement mechanisms to constrain the user's actions to those allowed by local policy. This approach removes the fine-grained policy configuration and decision making from the GT services (e.g., GridFTP, GRAM) and allows the local operating system to act as a sandbox. Thus, administrators can use normal policy administration tools to configure policy. For example, a Globus Toolkit user is normally mapped to a local Unix account. Standard Unix filesystem permissions, quotes, group memberships, and so forth are then used to configure and enforce policy.

Similar techniques could be used in conjunction with dynamically allocated accounts or virtual machine technology [9].

### 2.2.2. Limitations of GT Classic

The classic Globus Toolkit authorization system described has the advantage of being easy for site administrators to understand and configure because it uses existing local policy management and enforcement mechanisms with which the administrator is presumably already familiar. In terms of supporting a large VO, however, the GT has several shortcomings:

- *Scalability*: each personnel or policy change requires changing policy at each participating site;
- *Lack of expressiveness*: native OS methods may not be expressive enough to support VO policies;
- *Consistency*: different native OS methods may not support the same kinds of policies;
- *Distribution:* in order to maintain a consistent policy across the VO, each policy change must be propagated to each site involved. Any failure in propagation will cause an inconsistency in the policy.

To solve these problems, we undertook the development of the CAS system, described in the following section.

## 3. CAS CONCEPTS

We now introduce CAS and describe how it solves the problems described in Section 2.

### 3.1. Policy Management

CAS allows a VO to maintain its own set of policies explicitly and communicate those policies to sites. The sites then combine their local policies (about what the VO is allowed to do) with the VO's policies (about what the individual user is allowed to do as a VO member) and enforce this combined policy.

The VO, through administration of the CAS server, maintains the VO's portion of this combined policy. This portion of the policy includes the following:

- The VO's access control policies regarding its resources: which rights are granted to which users (e.g., which users can read which files);
- The CAS server's own access control policies, such as who can delegate rights or maintain groups with the VO. These policies can be expressed at a fine-grained level (e.g., a user may be allowed to grant rights on only certain resources, or add users to only certain groups);
- The list of VO members.

The other part of this combined policy, the resource provider's policy regarding the VO, is maintained by the resource provider using the same native mechanisms used for non-VO users. For example, a site may create a local identity representing a VO and add local configuration mapping users presenting credentials from that VO's CAS server to that identity. The site would then use local mechanisms to set policy on the VO as a whole, for example, change file ownerships to allow the VO identity read and write access to a particular subset of the file system, or set file system quotas limiting the amount of space that the VO can use. A resource provider may use this mechanism to maintain policies for several VOs, each running its own CAS server.

## 3.2. Policy Enforcement

Resource providers participating in a VO with CAS will deploy CAS-enabled services (i.e., services modified to enforce the policy in the CAS credentials) onto resources they assign to the VO. A user wishing to access those resources first contacts the VO's CAS server and requests a CAS credential. The CAS server replies with a CAS credential that contains a policy statement of that user's rights, cryptographically signed by the CAS server.

When making a request to the resource, the user presents the CAS credential. Upon receiving the CAS credential, a CAS-enabled service takes several steps to enforce both VO and local policy:

- Verify the validity of the CAS credentials (e.g., signature, time period).
- Enforce the site's policies regarding the VO, using essentially the same method as an unmodified server. However, the identity used when enforcing the site's policies is the identity of the signer of the policy assertion (i.e., the VO's CAS server), not the identity of the individual user authenticating.
- Enforce the VO's policies regarding the user, as expressed in the signed policy statement in the CAS credential.
- Optionally, enforce any additional site policies in regard to the user (for example, a site may keep a blacklist of end users who are not allowed to perform any action, regardless of any VO policy).

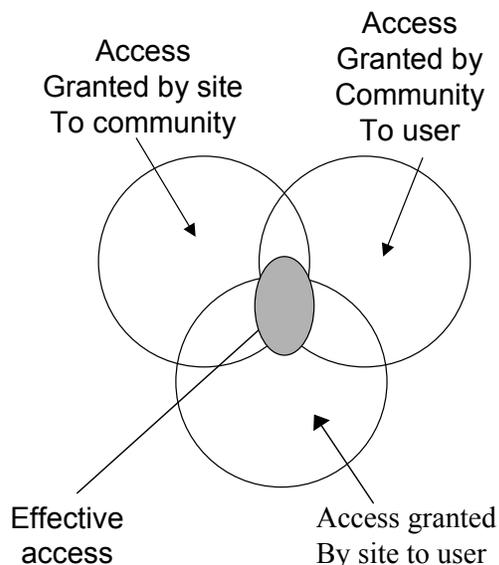

Figure 1: Effective access control policy in CAS.

Thus, as shown in Figure 1, the set of rights the user is effectively granted by these steps is the intersection of the set of rights granted by the resource server to the VO and the set of rights granted by the VO to the user (optionally modified by any site policy specific to the user).

## 4. CAS IMPLEMENTATIONS

We review past and current CAS implementations and discuss future implementation plans. We start with a review of Globus Toolkit proxy certificates, a key component of CAS implementations. Next we discuss the initial and current CAS prototype releases, explaining the differences between them and the motivations for those changes. We then discuss the upcoming CAS release in a future release of version 3 of the Globus Toolkit.

### 4.1. Proxy Certificate Overview

CAS is designed and implemented to work with the Grid Security Infrastructure (GSI) [2][13], which provides the security functionality of the Globus Toolkit. For authentication, GSI uses X.509 proxy certificates [22] to provide credentials for users and to allow for delegation and single sign-on.

Proxy certificates are similar to the standard X.509 end entity certificates (EECs) [6] from which they are derived. The primary difference is that users issue proxy certificates to create a short-term (e.g., one-day) delegation of their rights to another entity (e.g., a process running on their behalf) as opposed to EECs, which are issued by certificate authorities to assign a long-term identity. The short-term nature of proxy certificate credentials allows them to be more lightly protected than the credentials associated with the long-term EEC credentials that are used to create them. For example, proxy certificate credentials are usually protected with local filesystem permissions as opposed to being encrypted. This approach allows their use for single sign-on and by unattended processes.

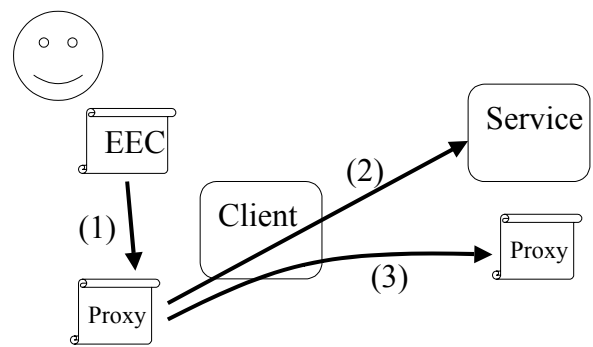

Figure 2: Proxy certificate creation and delegation. The steps are explained in the text.

As shown in Figure 2, step 1, the creation of a proxy certification can be local to a single resource (usually for use by subsequent clients, to enable single sign-on). As shown in step 2, the proxy certificate can then be used with GSI to authenticate to a remote service and establish a secure connection. Proxy certificates can also be used to issue other proxy certificates, allowing for a chain of delegations as shown in step 3, where a proxy certificate is created across the GSI-secure network channel (to delegate rights to the service so it can act on the original user's behalf).

Normally a proxy certificate allows its bearer to assert the full rights of the bear of the EEC that issued it. This delegation can be restricted through the use of a policy embedded in the proxy certification, as we discuss in the CAS implementation in the following section.

## 4.2. Initial CAS Prototype

Our initial CAS prototype [4], as described in [17], was released in March 2002. This implementation includes a CAS server, appropriate administration and user clients, and a GridFTP server [1] modified to understand and honor CAS credentials. The implementation uses the pyGlobus toolkit [16] to facilitate implementation using Python.

This implementation uses restricted proxy certificates, as described in Section 4.1, issued by the CAS server to the user. The proxy certificate identifies the user as a member of the VO (by virtue of the user's bearing a proxy certificate issued by the CAS server) and the rights the user possesses in the VO (through the restrictions in the proxy certificate). As shown in Figure 3, the effective rights of the user then become the subset of the rights granted by the resource to the VO and the VO to the user.

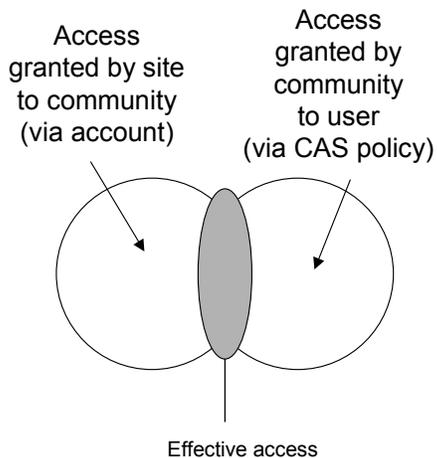

Figure 3: Effective access control policy in initial CAS prototype.

As described in Section 3.2, resources create an account for the virtual organization (VO) and grant some set of rights to the account. Each VO then sets up a CAS server and configures its policy in that server to express each user's rights within the VO.

Figure 4 shows the details of CAS alphaR1 use. The steps are as follows:

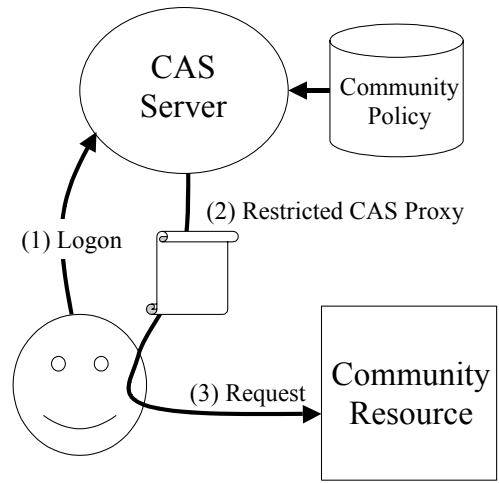

Figure 4: Use of initial CAS prototype. Steps are described in the text.

1. The user authenticates, using personal proxy credentials, to the VO's CAS server. The CAS authenticates the user and establishes that user's rights in the VO by using a local policy database.
2. The CAS server then issues to the user a restricted proxy certificate. This proxy certificate, as shown in Figure 5, has the identity of the CAS server, identifying the user as a VO member, and a restriction policy expressing the rights of the user within the VO.
3. The user uses the CAS proxy certificate to authenticate to a resource as a VO member. Based on that certificate, the resource allows the user to access the resource using the VO account but constrains the user's activities using the restrictions in the proxy certificate.

After the release of the initial CAS prototype in early 2002, we received useful feedback from both VO administrators and resource sites. The largest concern raised by the sites was their inability to identify the user connecting to them with CAS credentials at a finer level of granularity than a member of a particular VO (i.e., they could not verify the actual user identity). A number of these sites have policy requirements dictated by their funding agencies to identify users of their computational resources.

We proposed that the CAS server, which authenticates the user and hence has access to their identity, could encode the user identity in a non-critical X.509 extension [6] to the proxy certificate. This solution was also of concern to the resource providers because it required them to trust the CAS server as to the identity of the user. This caused them to place more trust in the CAS server than they were comfortable with.

This concern led to the redesign of CAS and the release of the alphaR2 prototype described in the following section.

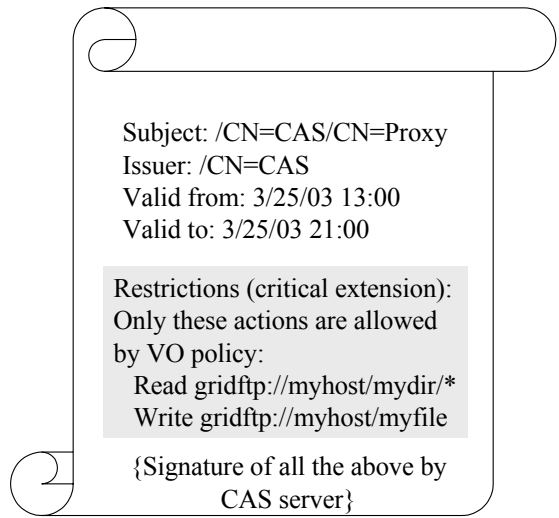

Figure 5: Example of a restricted proxy certificate, simplified for clarity, issued by initial CAS prototype. The certificate identifies the user as a member of the VO and the restrictions enumerate the user's rights within the VO.

### 4.3. Second CAS Prototype

Building on the first CAS prototype and the community feedback it elicited, we redesigned CAS and, in September 2002, released the alphaR2 prototype [5]. The major feature added in this release was to enable resources to identify a user operating with CAS credentials to the same degree of certainty as they could a user using standard Globus Toolkit proxy credentials. That is, a resource could derive not only the user's VO membership but also the user's identity. The contents of this release—a CAS server, clients and a CAS-enabled GridFTP server—are otherwise functionally identical to the first prototype.

To implement this feature, we changed the model from the use of a restricted proxy certificate issued by the CAS server to a model that combines a proxy certificate issued by the user with a signed policy assertion issued by the CAS server. The resulting credential is shown in Figure 6.

The policy assertions issued by the CAS server includes a set of access rights along with the user's identity. This assertion allows a resource to verify the user's VO membership, by the fact the assertion has the user's identity in it, and the user's rights within the VO as listed in the assertion. The assertion also includes a validity period and signature allowing resources to verify the validity of the assertion.

This separation of CAS policy assertion from user proxy credentials allows resources accepting these credentials to identify the user involved in a request as if they were using normal GSI credentials. This strategy in turn allows a site to use normal mechanisms for auditing and to apply an additional level of policy enforcement based on the user's identity. The applied policy, as shown in Figure 3, becomes the intersection of the rights granted to the community by the site, the rights granted to the user by the community and any restrictions placed on the user by the site.

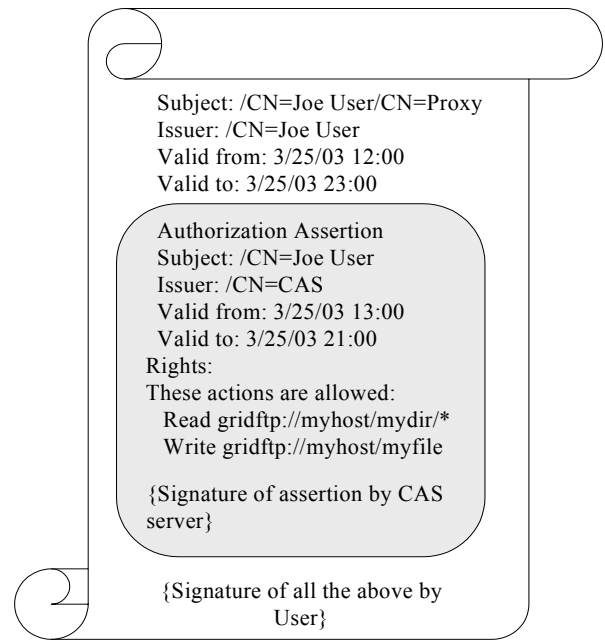

Figure 6: An example CAS alphaR2 credential, simplified for clarity. A user proxy is combined with a policy assertion issued by the CAS server.

Figure 7 shows the details of second CAS prototype usage. The steps in the figure are as follows:

1. The user authenticates, using personal proxy credentials, to the CAS server serving the user's virtual organization (VO). The CAS server established the user's identity and rights in the VO using a local policy database.
2. The CAS server issues the user a signed policy assertion containing the users identity and rights in the VO. The CAS client generates a new proxy certificate for the user and embeds the CAS policy assertion in the proxy certificate as a noncritical X.509 extension [6]. This embedding of the assertion in the proxy is not necessary from a security perspective but instead allows any application that can use a normal GSI proxy to be able to use the proxy with CAS assertion embedded in it, as it looks like a standard proxy certificate to existing APIs and protocols.
3. The user uses the proxy certificate with the embedded CAS assertion to authenticate to a resource. The resource authenticates the user using normal GSI authentication, allowing it to determine the user's identity and apply local policy as desired. It then parses the policy assertion in the proxy certificate and, based on the assertion, allows the user to access the resource using the VO account but constrains the user's activities using the rights in the assertion.

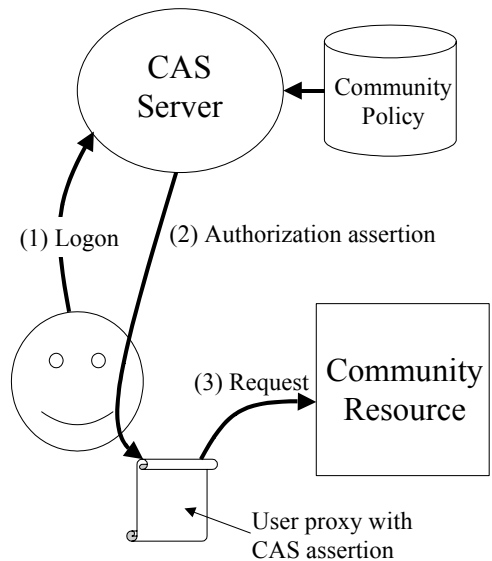

Figure 7: Second CAS prototype usage. Steps are described in the text.

### 4.4. Summer 2003 Release

Building from the earlier prototypes described in the preceding sections, we are planning on releasing a new version of CAS built on version 3 of the Globus Toolkit (GT3) [15]. This version of CAS will be conceptually similar to the second prototype described in the preceding section but will have a number of significant differences:

- *Implements an OGSA Service.* It will reside in a GT3 hosting environment and implement an Open Grid Service Architecture [11] service. Protocols to access CAS will use common Web services methods such as SOAP [20].
- *Has a Java code base.* To facilitate its integration with GT3, the CAS will be written in Java.
- *Uses SAML for assertion format.* The new version of CAS will use the security assertion markup language (SAML) [19] as the format for the policy assertions issued by the CAS server. This will allow for easy integration with Web services and OGSA tooling.

Our intent is to release this new version of CAS with a version of the Globus Toolkit in late summer 2003. This release will include many functional elements present in the earlier prototypes, namely, a CAS server, clients for users and administrators, and a CAS-aware GridFTP server. It will also include a Java client library and a C library to allow for CAS-enabling services (i.e., allowing them to accept and properly enforce CAS-issued assertions).

**TUBT003**

## 5. RELATIONSHIP TO OTHER WORK

We compare CAS to other authorization services and explain how we see standardization allowing these services to interoperate.

### 5.1. VOMS

The Virtual Organization Management Service (VOMS) [23] and CAS are similar architecturally in that both issue policy assertions to a user that the user then presents to a resource for the purpose of obtaining VO-issued rights. The primary difference between the two systems is the level of granularity at which they operate.

As shown in Figure 8, VOMS assertions contain a list of role or group memberships held by the user. The user presents this assertion to a resource, which then determines that user's rights based on their memberships and local policies about those memberships. In effect the policy about what memberships a user has is centralized in the VOMS server, but the policy regarding exactly what rights those memberships grant is distributed among the sites.

CAS assertions provide the rights directly and do not need interpretation by the resource. As discussed in Section 2, in situations where policies are changing dynamically, we believe this complete centralization of policy can achieve better consistency.

However, while CAS was designed primarily to do fine-grained policies, our research has shown it is also capable of asserting coarser grained group memberships [3].

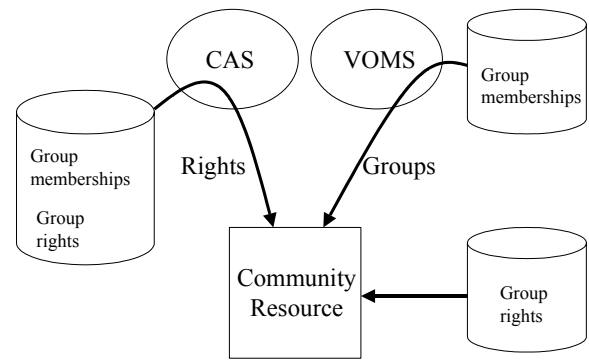

Figure 8: Difference between CAS and VOMS models. VOMS issues group memberships, which are mapped to rights by a resource. CAS issues rights directly.

### 5.2. Akenti and PERMIS

Akenti [21] and PERMIS [7], while having differences in implementation and features, are architecturally similar in that they provide a resource with an authorization decision in regards to a request. Both follow the basic model:

1. A resource authenticates a requestor and validates their identity as well as possibly some additional attributes.
2. The resource receives and parses the user's request.
3. The resource then passes the identity, attributes, and request to Akenti or PERMIS and requests an authorization decision (i.e., whether is should service or reject the request).
4. Akenti or PERMIS then returns a decision to the resource and which enforces it.

While our CAS implementations provide simple authorization decision functionality, they are limited to supporting CAS policy assertions and does not have as rich a feature set as either Akenti or PERMIS. It is possible that either of these systems, with some modifications, could be used to provide resource-side functionality for CAS (i.e., parse the CAS assertion and use it to authorize the user's request.)

### 5.3. Standardization of Assertions

We see potential for allowing interoperability between CAS and the other authorization systems mentioned in this section by standardizing on a format for authorization assertions. Today, both CAS and VOMS each issue assertions in a different and nonstandard format that requires custom code on the resource side to interpret for authorization.

By standardizing the format for the assertions issued by services such as CAS and VOMS, we can enable authorization decision services to be able to parse and use these assertions in their decision-making as shown in Figure 9. The steps are as follows:

1. An assertion service like CAS or VOMS would issue an assertion to a user regarding either that user's attributes (e.g., group memberships) or rights.
2. The user would present this assertion to a resource when making a request (probably along with authenticating their identity.)
3. The resource would then present the request, identity and assertion to an authorization decision service such as Akenti or PERMIS.
4. The authorization decision service would parse the assertion and use it, along with local policy, to render a decision that it returns to the resource.

X.509 attribute certificates [6] are one possibility for a standard assertion format. Given the influence of Web services in emerging Grid standards [11], however, we believe a standard more in line with current Web services efforts should be considered. We are experimenting with the security assertions markup language (SAML) [19], a standard for formatting authorization and attribute assertions, as a format for our next release of CAS as described in Section 4.4. We also plan on working in the Global Grid Forum [14] in order to standardize the use of SAML for supporting authorization in VOs.

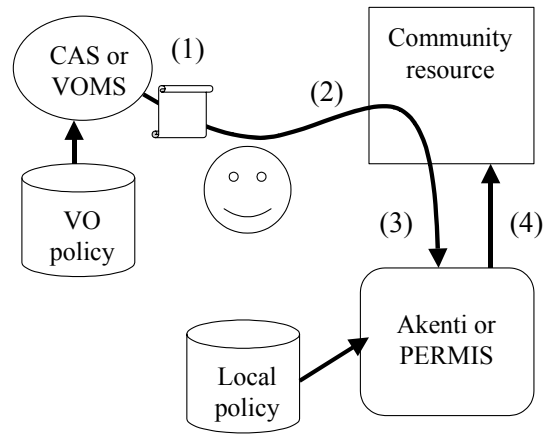

Figure 9: A standard assertion format would allow authorization systems such as CAS, VOMS, Akenti and PERMIS to interoperate. Steps are described in the text.

## 6. FUTURE WORK

We plan future research activities in regard to CAS.

### 6.1. Caching Server

The current CAS server implementation maintains a single, centralized server for a VO. To provide for availability even if the CAS server is unavailable, we will develop a caching server to act as a lightweight partial mirror of a CAS server. This caching server will accept requests to cache certain types of signed policy statements (for example, all rights granted to a user) and then periodically request these signed statements from the VO's CAS server. The caching server will accept CAS-protocol requests from users by returning these cached signed policy statements.

### 6.2. Resource-Server-Pull Support

In addition to current model in which the user pushes the policy assertions to the resource server, we will investigate a model in which the resource server, rather than the client, contacts the CAS server for policy assertions.. This model could be combined with the caching server, described in the preceding section, for performance and reliability: the caching server could periodically contact the VO's CAS server for a signed policy assertion including all rights granted on a resource, and the resource server could query that caching server.

### 6.3. Local Authorization Server

We are investigating the possibility of implementing a local authorization server that would accept authorization queries from request servers, apply all applicable local and community policies, and return a yes or no answer. This authorization server would need to be highly trusted by the resource server and highly available (sites would probably run a local authorization server on each resource server host.)

This service could potentially take CAS credentials, forwarded by the resource, and use their credentials in making its decision, or it could contact the CAS server itself using the pull model described in the preceding section.

Such a server could be implemented by using Akenti or PERMIS, as described in Section 5

## Acknowledgments

We thank Rachana Ananthakrishnan, Doug Engert, Sam Meder, Chris Nebergall, and Shubi Raghunathan for their contributions to the Community Authorization Service.

We gratefully acknowledge feedback on the CAS implementations from members of the PPDG-SiteAAA and DOE Science Grid engineering groups and Keith Jackson for the pyGlobus [16] system used in our prototypes.

This work was supported in part by the Mathematical, Information, and Computational Sciences Division subprogram of the Office of Advanced Scientific Computing Research, Office of Science, SciDAC Program, U.S. Department of Energy, under Contract W-31-109-ENG-38. Initial funding for CAS was provided by the "Earth Systems Grid" project.

The Globus Toolkit is a trademark owned by the University of Chicago.

## References

[1] W. Allcock, J. Bester, J. Bresnahan, A. L. Chervenak, I. Foster, C. Kesselman, S. Meder, V. Nefedova, D. Quesnal, and S. Tuecke. Data Management and Transfer in High Performance Computational Grid Environments. *Parallel Computing Journal* 28 (5), May 2002, pp. 749-771.

[2] Butler, R., Engert, D., Foster, I., Kesselman, C., Tuecke, S., Volmer, J. and Welch, V. A National-Scale Authentication Infrastructure. *IEEE Computer*, *33* (12), 2000, pp. 60-66.

[3] S. Cannon, S.Chan, D.Olson, C. Tull, V. Welch, L. Pearlman. Usering CAS to Manage Role-ased VO Sub-Groups. To appear CHEP 03, June 2003.

[4] CAS Alpha Release Web site, http://www.globus.org/Security/cas/alpha/, March 2002.

[5] CAS AlpahR2 Web site, http://www.globus.org/Security/cas/alpha-r2/, September 2002.

[6] CCITT Recommendation X.509: The Directory – Authentication Framework. 1988.

[7] D. W. Chadwick and A. Otenko. The PERMIS X.509 Role Based Privilege Management Infrastructure. *7th ACM Symposium on Access Control Models and Technologies*, 2002.

[8] S. Farrell, and R. Housley. An Internet Attribute Certificate Profile for Authorization, RFC 3281, IETF, April 2002.

[9] R. Figueiredo, P. Dinda, and J. A. Fortes. Case for Grid Computing on Virtual Machines. *23rd International Conference on Distributed Computing Systems*, 2003.

[10] I. Foster and C. Kesselman. Globus: A Metacomputing Infrastructure Toolkit. *International Journal of Supercomputer Applications*, 11 (2). 115-129. 1998

[11] I. Foster, C. Kesselman, J. Nick, and S. Tuecke. The Physiology of the Grid: An Open Grid Services Architecture for Distributed Systems Integration, Globus Project, 2002.

[12] I. Foster, C. Kesselman, and S. Tuecke. The Anatomy of the Grid: Enabling Scalable Virtual Organizations. *International Journal of High Performance Computing Application,* 15 (3), 2001, pp. 200-222.

[13] I. Foster, C. Kesselman, G. Tsudik,. and S. Tuecke. A Security Architecture for Computational Grids. *ACM Conference on Computers and Security*, 1998, pp. 83-91.

[14] The Global Grid Forum, www.ggf.org, May 2003.

[15] The Globus Toolkit 3.0 Alpha Release, 2003 http://www.globus.org/ogsa/releases/alpha/index.html

[16] Jackson, K. pyGlobus: a Python Interface to the Globus Toolkit. *Concurrency and Computation: Practice and Experience*, *14* (13-15), 2002, pp. 1075-1084.

[17] L. Pearlman, V. Welch, I. Foster, C. Kesselman, and S. Tuecke. A Community Authorization Service for Group Collaboration. *IEEE 3rd International Workshop on Policies for Distributed Systems and Networks*, 2002.

[18] D. Reed, I. Pratt, P. Menage, S. Early, and N. Stratford. Xenoservers: Accountable Execution of Untrusted Programs. *7th Workshop on Hot Topics in Operating Systems*, Rio Rico, AZ, IEEE Computer Society Press, 1999.

[19] Security Assertion Markup Language (SAML) 1.0 Specification, OASIS, November 2002.

[20] Simple Object Access Protocol (SOAP) 1.1, W3C, 2000.

[21] M. Thompson, W. Johnston, S. Mudumbai,.G. Hoo, K. Jackson,. and A. Essiari. Certificate-based Access Control for Widely Distributed Resources. *8th Usenix Security Symposium*, 1999.

[22] S. Tuecke, D. Engert., I. Foster, V. Welch, M. Thompson, L. Pearlman, and C. Kesselman. Internet X.509 Public Key Infrastructure Proxy Certificate Profile, IETF, 2003.

[23] VOMS Architecture v1.1, http://grid-auth.infn.it/docs/VOMS-v1_1.pdf, May 2002.